\documentclass[preprint,12pt]{elsarticle}

\usepackage{amsmath,amssymb}
 \usepackage{color}
 \usepackage{ulem}

\begin{document}

\begin{frontmatter}



\title{Skyrmions on 2D Elastic Surfaces with Fixed Boundary Frame}


\author{Sahbi EL Hog$^1$, Fumitake Kato$^{2}$ }
\author{Hiroshi Koibuchi$^{3}$ }
\ead{koibuchih@gmail.com}
\author{Hung T. Diep$^1$}
\ead{diep@u-cergy.fr} 
\address{  $^1$Laboratoire de Physique The${\acute o}$rique et Mod${\acute e}$lisation, University of Cergy-Pontoise, CNRS, UMR 8089 2, Avenue Adolphe Chauvin, 95302 Cergy-Pontoise Cedex, France \\
$^2$Department of Industrial Engineering, National Institute of Technology (KOSEN), Ibaraki College, Nakane 866, Hitachinaka, Ibaraki 312-8508, Japan\\
$^3$Department of General Engineering, National Institute of Technology (KOSEN), Sendai College, 8 Nodayama, Medeshima-Shiote, Natori-shi, Miyagi 981-1239, Japan
}


\begin{abstract}
We report simulation results of skyrmions on fluctuating 2D lattices, where the vertices ${\bf r}_i (\in {\bf R}^3)$ are treated as a dynamical variable  and, hence, there is no crystalline structure. On the fluctuating surfaces,  an external magnetic field perpendicular to the surface, Dzyaloshinskii-Moriya and ferromagnetic interactions are assumed in addition to the Helfrich-Polyakov Hamiltonian for membranes.  The surface (or frame) tension $\tau$ is calculated  under both isotropic and uniaxial strain conditions, and this calculation clarifies a non-trivial dependence of $\tau$ on the skyrmion, stripe, and ferromagnetic phases. We find that the variation of $\tau$ with respect to the applied magnetic field in the skyrmion phase is accompanied by a variation of the total number of skyrmions. Moreover, we find that this total number variation is qualitatively consistent with a recent experimental result for the creation/annihilation of skyrmions of 3D crystalline material under uniaxial stress conditions.  It is also found that the stripe phase is significantly influenced by uniaxial strains, while the skyrmion phase remains unchanged.  These results allow us to conclude that the skyrmion phase is stable even on fluctuating surfaces.
\end{abstract}




\end{frontmatter}


\section{Introduction\label{intro}}
It is widely accepted that skyrmion configurations, which are observed on materials such as FeGe, MnSi, or ${\rm Cu_2OSeO_3}$, is a promising candidate for future computer memories because of their dramatic energy-saving property in electric transportation.  Due to the topological nature of the spin configuration, its stability against external stimuli such as thermal fluctuations is one of its most interesting properties \cite{Skyrme-1961,Dzyalo-1964,Moriya-1960,Bogdanov-Nat2006,Bogdanov-PHYSB2005,Bogdanov-SovJETP1989}. For these reasons, many experimental and theoretical studies have been conducted by assuming competing interactions, such as ferromagnetic (FM) interaction, Dzyalonskii-Moriya (DM), and Zeeman energy under constant magnetic fields \cite{Munzer-etal-PRB2010,Mohlbauer-etal-Science2009,Buhrandt-PRB2013,Dong-etal-PRB2010,Zhou-Ezawa-NatCom2014,Iwasaki-etal-NatCom2013}. In particular, skyrmions in 2D systems have attracted a great deal of attention due to their potential applications in memory devices \cite{Bogdanov-ZETP1989,Fukuda-Z-Natcom2011,Banerjee-etal-PRX2014,Gungordu-etal-PRB2016}. However, the creation/annihilation of skyrmions and the stability of the skyrmion phase are yet to be studied in the context of such applications.

Along this research direction, several investigations have been conducted on the responses to mechanical stresses, which  have a non-trivial effect on the spin configuration \cite{Nakajima-etal-PRB2011,Ritz-etal-PRB2013}.  In Ref.\cite{Shi-Wang-PRB2018}, Shi and Wang numerically find that skyrmions are stabilized due to the magnetoelastic coupling even without
external magnetic field by introducing an explicit magnetoelastic interaction
energy.  This confirms that skyrmions may be generated by various mechanisms.  We come back to this point later in next section.
 Nii et al. recently reported that MnSi is an anisotropic elastic object \cite{Nii-etal-PRL2014}, and they also reported the experimental results of a stress-induced skyrmion-to-conical phase transition in MnSi, where the applied uniaxial stresses create/annihilate skyrmions under suitable external magnetic fields \cite{Nii-etal-NatCom2015}. It is also reported by Seki et al. that skyrmions in ${\rm Cu_2OSeO_3}$ can be stabilized by a uniaxial tensile strain  \cite{Seki-etal-SCI2012,Seki-etal-PRB2017}, and Levatic et al. reported that mechanical pressure significantly increases the skyrmion pocket  in ${\rm Cu_2OSeO_3}$ \cite{Levatic-etal-SCRep2016} and in  MnSi \cite{Chacon-etal-PRL2015}. From these experiments, it was clarified that uniaxial mechanical stimuli stabilize/destabilize the skyrmion phase, or, in other words, a small lattice deformation enhances the stability/instability of the skyrmion phase.  This indicates that the interplay between two different degrees of freedom spin and lattice, which can be rephrased by spin and orbit, plays an important role in mechanical responses, which has been extensively studied via DM interaction energy. Chen et al. numerically studied skyrymions in MnSi by using a 3D model and suggested that  anisotropies in DM and FM interactions play an important role in their stability under uniaxial stress in a wide range of temperatures including low temperature region \cite{Chen-etal-SCRep2017}. These anisotropies are naturally expected when crystalline lattices are considerably deformed. Thus, it is interesting to study the skyrmion stability under lattice deformations at finite temperatures. 

The purpose of this paper is to see whether the skyrmion phase is stable under large lattice deformations. For this purpose, it is interesting to assume the lattice degrees of freedom as a dynamical variable. Here we should note that "dynamical" variable refers to the variable that is integrated in the partition function in the context of statistical mechanics. Therefore, a fluctuating lattice is considered to be a direct tool for this purpose, where the lattice fluctuation is allowed not only into the in-plane direction but also into the out-of-plane direction  \cite{HELFRICH-1973,POLYAKOV-NPB1986,WIESE-PTCP19-2000,NELSON-SMMS2004,GOMPPER-KROLL-SMMS2004,KANTOR-NELSON-PRA1987,Essa-Kow-Mouh-PRE2014,Noguchi-JPS2009}. On the fluctuating lattices,
a certain magnetoelastic coupling is implicitly included in the DM interaction, and for simplicity, only this magnetoelastic coupling is taken into account under the presence of external magnetic field in this paper. Long-range magnetic dipole-dipole interaction is neglected and no anisotropy is assumed in the FM and DM interactions. 
Skyrmions on fluctuating lattice are interesting also from the view point of liquid crystal skyrmions \cite{Fukuda-Z-Natcom2011,Bogdanov-PRE2003,Bogdanov-PRE2014}.

The DM interaction on the continuous 2D plane depends on both spins and coordinates, and hence, we are able to calculate mechanical properties such as surface tension under the presence of skyrmions if the coordinate, or equivalently the surface, is treated as a dynamical variable \cite{HELFRICH-1973,POLYAKOV-NPB1986,WIESE-PTCP19-2000,NELSON-SMMS2004,GOMPPER-KROLL-SMMS2004,KANTOR-NELSON-PRA1987,Essa-Kow-Mouh-PRE2014,Noguchi-JPS2009}.  Moreover, DM interaction is invariant under an arbitrary rotation around the axis perpendicular to the surface. However, its discrete version, which will be described in the next section, has only a discrete symmetry on 2D rigid lattices because the rigid lattices have only discrete symmetries. In contrast, the fluctuating lattices, where the edge direction of the lattice becomes almost random, are not regarded as a rigid lattice by any rotation. For this reason, the discrete DM interaction is expected to be influenced by the surface fluctuations. 

Therefore, it is worth while studying skyrmions to see the interplay between spins and surface elasticity on fluctuating surfaces. In this paper, we calculate the surface tension $\tau$ by Monte Carlo (MC) simulations to study the influences of skyrmion configurations on $\tau$. Three different spin configurations are expected on the surface; stripe, skyrmion, and FM phases, as in the case of rigid lattices \cite{Hog-etal-JMagMat2018,Yu-etal-Nature2010,Yu-etal-PNAS2012,Kwon-etal-JMagMat2012}. Phase diagrams are also obtained on a plane with axes of temperature and external magnetic field. In this calculation, 

A mechanical stress can be applied to the elastic lattices only in the form of strain, which is given by fixing the boundary to a suitable size or shape. Such an effective stress becomes isotropic in general, because the lattice structure changes relatively freely due to the fact that the vertex position is a dynamical variable. Nevertheless, the uniaxial stress condition is simulated by modifying the boundary lengths of the rectangular lattice from its isotropic length such as from $(L_1,L_2)$ to $(L_1/\xi, \xi L_2), (\xi\!\not=\!1)$, for example. As a result of this geometry modification, it will be checked whether the stripe domain axis aligns along the uniaxial stress direction. We also examine whether this uniaxial stress condition influences the stability of skyrmion phase under an external magnetic field.

\section{Triangulated lattices and skyrmion model\label{lattice-model}}
\subsection{Triangulated lattices\label{lattice}}
\begin{figure}[t]
\begin{center}
\includegraphics[width=8.5cm]{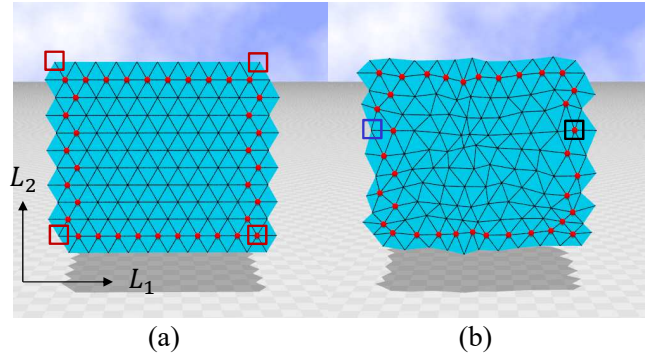}
 \caption{ (Color online) (a) The initial configuration of triangulated lattices of size $(L_1,L_2)=(10,10)$, where $L_1$ and $L_2$ are the total number of vertices along the horizontal ($L_1$) and vertical ($L_2$) directions, respectively, and (b) a snapshot of the lattice after a sufficiently large number of Monte Carlo updates. Periodic boundary condition  (PBC) is assumed in both directions. The BVs are denoted by small spheres, outside of which represents the virtual vertices connected by the PBC. The vertices enclosed by squares without the small sphere are virtual points corresponding to the real vertex with a small sphere in (a) and (b). }
 \label{fig-1}
\end{center}
\end{figure}
A triangulated lattice with a square boundary is used for the simulations. Triangular lattices have already been used to study skyrmions using non-DM \cite{Okubo-etal-PRL2012} and DM interaction energies \cite{Rosales-etal-PRB2015}. The total number of vertices $N$ is given by $N\!=\!L^2$, where $L$ is the number of vertices on one boundary edge. In Fig. \ref{fig-1}(a), an initial configuration of the lattice of $N\!=\!100 (\Leftrightarrow L\!=\!10)$ is shown, where the boundary vertices (BVs) are denoted by small spheres.  The lattice size $N\!=\!100$ is small enough to visualize the lattice structure including the BVs, and it is  $10^2$ times smaller than the size of lattices used in the simulations.  Let $a$ be the edge length of the initial regular triangle, which forms the initial lattice in Fig. \ref{fig-1}(a), then the side lengths along the horizontal ($L_1$) and vertical ($L_2$) directions are 
\begin{eqnarray}
\label{latt-size}
\left((L-1)a, (L-1)a\sqrt{3}/2\right), \quad a=1. 
\end{eqnarray}
  We call this $a$ the lattice spacing  \cite{Creutz-txt} of the undeformed lattice shown in  Fig. \ref{fig-1}(a). Note that each bond length changes due to the surface fluctuations when elasticity is turned on. 

 If the simulation results such as $\tau$ are compared with experimental data, this $a$ can be fixed to a suitable value \cite{Koibuchi-PRE2017}. In this paper, we are interested in the response of $\tau$ to  isotropic and uniaxial strains under the variation of external magnetic field and temperature, and therefore we fix $a$ to $a\!=\!1$, $a\!=\!0.75$, and $\xi^{\pm 1}a$ ($\xi^2\!=\!0.9, 1.1$ for $a\!=\!1$), which are direction dependent, in the simulations.

The vertices inside the boundary are allowed to move three-dimensionally, while the BVs can move only in the 2D plane parallel to the   boundary square with the periodic boundary condition (PBC), or in other words, the BVs are moving into the in-plane directions under the condition that the vertices are connected to the opposite BVs, which are placed at the positions $\pm L_{1,2}$ from the original positions. To describe this periodic boundary condition in more detail, we plot in Fig. \ref{fig-1}(b) a snapshot of a typical configuration after a sufficiently large number of Monte Carlo (MC) updates, which will be described below. 
 In these Figs. \ref{fig-1}(a) and \ref{fig-1}(b), the vertices outside the BVs are the opposite BVs. Thus, the lattices used in the simulations are summarized as follows: (i) the BVs move only in the plane parallel to the boundary square, (ii) the PBCs are imposed on the BVs. 
 
We should comment on the reason why the above mentioned PBC is imposed on the BVs. First of all, the boundary condition imposed on the surfaces is necessary for the calculation of the surface tension $\tau$, which will be described below in detail. For this reason, the PBC assumed here is slightly different from the ordinary PBC, which is simply introduced to lower the boundary effects. Moreover, the PBC assumed here on the lattice  should be connected or close to experimental ones, where the $z$ direction is mechanically fixed at the boundary. For this reason, it is reasonable to assume that the BVs are prohibited at least from moving in the $z$ direction. It is also possible to assume that the BVs are completely fixed for the evaluation of $\tau$. Hence, the problem is whether the results depend on the boundary conditions or not. Although the BC that the BVs are completely fixed is not examined, we expect that the results are independent of these BCs, because the BVs move only slightly as we will see in the snapshots of lattices below. 

Let us define the area $A_p$ of the boundary frame without deformation by elastic effects, by using $L_1$, $L_2$ and $a$,
\begin{eqnarray}
\label{projected_area}
A_p=(L_1-1)(L_2-1)\sqrt{3}a^2/2,\quad  a=1.
\end{eqnarray}
Under the elastic effect, the membrane fluctuates out of the $xy$ plane. However, we see that $A_p$ is identical with the projected membrane area if the BVs are completely fixed on the boundary frame.  $A_p$ is not identical to the area of the projected membrane if the BVs are allowed to move in the $xy$ plane at $z\!=\!0$ (see Fig. 1(b)). This boundary condition is mechanically imposed on materials to pin the membrane boundary in the $z\!=\!0$ plane so that
fluctuations in the $z$ direction do not drift the whole system in the space during the simulation (see Fig. 2a for illustration). Such a mechanically pinning BC does not restraint microscopic in-plane movements of particles.

\subsection{Skyrmion model\label{Sky-model}}
We start with a continuous version of DM interaction \cite{Dzyalo-1964,Moriya-1960,Bogdanov-Nat2006}, which is given by
\begin{eqnarray}
\label{cont_DM}
S_{\rm DM}=\int d^2x \frac{\partial {\bf r}}{\partial x_a}\cdot \left( \sigma\times \frac{\partial \sigma}{\partial x_a}\right),
\end{eqnarray}
where $x_a (a=1,2)$ denotes a parameter or a local coordinate of the 2D plane, and ${\bf r} (\in {\bf R}^3)$ is the position vector of the plane. It is clear from this expression that the surface shape and the spin are interacting to each other, though this form is sufficiently simple for the DM interaction. Note also that this expression of $S_{\rm DM}$ is identical with $S_{\rm DM}^\prime=\int d^2x \sigma\cdot \left( \nabla\times \sigma\right)$ if ${\partial {\bf r}}/{\partial x_a}$ in  $S_{\rm DM}$ is regarded as the unit tangential vector along $x_a$ direction.  
 We should note that  $\partial_a {\bf r}$ and $\sigma\times \partial_a \sigma$ in $S_{\rm DM}$ are vectors in ${\bf R}^3$, and the latter one is a pseudo vector, which remains unchanged under the inversion ${\bf r}\to -{\bf r}$ in  ${\bf R}^3$. Hence $\sum_{a} \partial_a {\bf r}\cdot \left(\sigma\times \partial_a \sigma\right)$ or the energy $S_{\rm DM}$ is a pseudo scalar in ${\bf R}^3$. For this reason, $S_{\rm DM}$ is invariant under an arbitrary rotation around the axis perpendicular to the lattice at least, because a pseudo scalar shares the same property with a scalar under space rotations. This allows us to choose a rigid lattice to define the discrete Hamiltonian, which will be introduced below. Real materials also have a crystalline structure.  In contrast, fluctuating lattices are not always connected to the crystalline lattice by any rotation. For this reason, it is interesting to study the skyrmion stability on fluctuating lattices. Moreover, an interaction of spins and lattices can be reflected in the frame tension if the lattice is treated as a dynamical variable as mentioned in the Introduction. 

To study skyrmions on the fluctuating lattices, we have to combine the skyrmion model and the elastic surface model \cite{WIESE-PTCP19-2000,NELSON-SMMS2004,GOMPPER-KROLL-SMMS2004,KANTOR-NELSON-PRA1987,Essa-Kow-Mouh-PRE2014,Noguchi-JPS2009}. If the two Hamiltonians are merged, the total Hamiltonian becomes slightly lengthy, but the definition itself is straightforward.   
Indeed, the Hamiltonian $S$ is given by a linear combination of several terms such as
\begin{eqnarray}
\label{model}
&&S({\bf r},\sigma)=S_1+\kappa S_2+ \lambda S_F + \delta S_{\rm DM} + S_B, \nonumber \\
&&S_1=\sum_{ij}\ell_{ij}^2, \quad S_2=\sum_{ij}\left(1-{\bf n}_i\cdot{\bf n}_j\right), \nonumber \\
&&S_F=\sum_{ij}\left(1-\sigma_i\cdot\sigma_j\right), \quad S_{\rm DM}=\sum_{ij}{\bf e}_{ij}\cdot \sigma_i\times \sigma_j,\\
&&S_B=\sum_i \sigma_i\cdot \vec B, \quad \vec B=(0,0,B).  \nonumber
\end{eqnarray} 
The unit of energy is given by $k_B\!=\!1$ (in $k_BT)$, which we call simulation unit, where $k_B$ and $T$ are the Boltzmann constant and the temperature, respectively.
The symbols ${\bf r},\sigma$ in $S$ denote the vertices and spins such that ${\bf r}\!=\!\{{\bf r}_1,{\bf r}_2,\cdots, {\bf r}_N\}$,  $\sigma\!=\!\{\sigma_1,\sigma_2,\cdots, \sigma_N\}$, where ${\bf r}_i\in {\bf R}^3$ (${\bf r}_i\in {\bf R}^2$ for BVs) and $\sigma_i\in S^2$(: unit sphere). The first and second terms are the discrete Hamiltonians of the surface model of Helfrich and Polyakov \cite{HELFRICH-1973,POLYAKOV-NPB1986}. These $S_1$ and $S_2$ are called Gaussian bond potential and bending energy, respectively, and $\ell_{ij}(=\!|{\bf r}_j\!-\!{\bf r}_i|)$ is the length of bond $ij$, and ${\bf n}_i$ and  ${\bf n}_j$ in $S_2$ are the unit normal vectors of the triangles that share the bond $ij$. The parameter $\kappa$ is the bending rigidity of the surface  \cite{WIESE-PTCP19-2000,NELSON-SMMS2004,GOMPPER-KROLL-SMMS2004,KANTOR-NELSON-PRA1987,Essa-Kow-Mouh-PRE2014}.  We should note that the first term $S_1$ depends on the size of projected area of lattice because $S_1$ is the sum of bond length squares of surface which spans the boundary of projected area $A_p$, and for this reason the model becomes dependent on the lattice spacing $a$. Indeed, for sufficiently large (small) $a$, fluctuations of the surface spanning the boundary are expected to be suppressed (enhanced). 
Here we should also emphasize that since the boundary is fixed during the simulations, $A_p$ is not changed with the surface fluctuations.  For this reason, the undeformed-lattice spacing a is also independent of the surface fluctuations as seen in Eq. (\ref{projected_area}).

The third term $\lambda S_F$ is the energy for describing the FM interaction between two neighboring spins with the interaction strength $\lambda$.  The fourth term $\delta S_{\rm DM}$ describes the DM interaction with the interaction coefficient $\delta$, where ${\bf e}_{ij}(=\!({\bf r}_j\!-\!{\bf r}_i)/|{\bf r}_j\!-\!{\bf r}_i|)$ is the unit tangential vector from ${\bf r}_i$ to ${\bf r}_j$. This vector ${\bf e}_{ij}$ corresponds to $\partial {\bf r}$ in the continuous $S_{\rm DM}$ in Eq. (\ref{cont_DM}) and varies on the fluctuating lattices. It is also possible to use $\vec{\ell}_{ij}(=\!{\bf r}_j\!-\!{\bf r}_i)$ for  $\partial {\bf r}$, however, here we use the unit length vector ${\bf e}_{ij}$ for $\partial {\bf r}$. 
The continuous expression $\sigma\times \partial_a \sigma$ can also be reduced to $\sigma_j\times\sigma_i$ in the discrete $S_{\rm DM}$, because the differential $\partial_a \sigma$ is replaced by the difference  $\sigma_j\!-\!\sigma_i$. 
 Thus, the term $S_{\rm DM}$ plays a role in the interaction between $\sigma$ and ${\bf r}$, the spins and the surface shape. The final term $S_B$ is the energy for the external magnetic field $\vec B\!=\!(0,0,B)$, which is perpendicular to the plane of the BVs. 
 
We note that effects of surface fluctuations on spins are taken into consideration only via the DM interaction energy $S_{\rm DM}$. The tangential vector ${\bf e}_{ij}$ in $S_{\rm DM}$ is deformed by  the surface fluctuations as mentioned above. On the other hand, the FM interaction can also reflect surface fluctuations, because the magnetoelastic coupling is not negligible on the fluctuating surfaces  due to the spin-orbit coupling \cite{Sander-RPP1999}. 
However, in the case of skyrmion deformations in a strained crystal, it is expected that the deformation of DM interaction plays an essential role, which is precisely checked in  \cite{Shibata-etal-Natnanotech2015}. Although the study in this paper is not for the skyrmion shape deformation, we simply use the ordinary undeformed $S_F$ in Eq. (\ref{model}). Thus, we neglect the effect of magnetoelastic coupling due to the FM interaction and study only effects of surface elasticity on spins via the DM interaction deformation. 

The partition function is given by 
\begin{eqnarray}
\label{part-func}
\begin{split}
&Z(\beta,\kappa,\lambda,\delta,B)=\sum_\sigma \int \prod_{i=1}^{N-N_{\rm BD}} d{\bf r}_i \int \prod_{i=1}^{N_{\rm BD}} d{\bf r}_i C(A_p)\exp\left[ -\beta S({\bf r},\sigma) \right],\\
& \beta=1/T, \quad (\Leftrightarrow  k_B=1)
\end{split}
\end{eqnarray} 
where $\sum_\sigma$ denotes the sum of all possible values of $\sigma$. The symbol $\beta$ is the inverse temperature, and the simulation unit is  defined by the relation $k_B\!=\!1$, and, hence, the temperature $T$ is not the real temperature. The symbols $\int \prod_i d{\bf r}_{i=1}^{N-N_{\rm BD}}$, $\int \prod_i d{\bf r}_{i=1}^{N_{\rm BD}}$ denote the three-dimensional and two-dimensional multiple integrations of the vertices inside and on the boundary, respectively, where  $N_{\rm BD}$ is the total number of vertices on the boundary. {The symbol $C(A_p)$ denotes a constraint on the integration of the boundary vertices so that the projected area remains fixed to $A_p$. This constraint is a function of ${\bf r}_1,\cdots, {\bf r}_{N_{\rm BD}}$, however this function $C({\bf r}_1,\cdots,{\bf r}_{N_{\rm BD}};A_p)$ is too complex to write down with $A_p$,  and for this reason we use $C(A_p)$ \cite{WHEATER-JP1994}. 

\subsection{Frame tension\label{frame-tension}}
As mentioned in the Introduction, only strains can be imposed on the elastic lattices by fixing the edge lengths of the rectangular boundary frame \cite{WHEATER-JP1994,Cai-Lub-PNelson-JFrance1994}. The imposed strains uniquely determine the corresponding stresses due to the stress-strain correspondence such as the stress-strain diagram, at least in the elastic deformation region specific to the materials. From this correspondence, stresses are effectively imposed on the elastic lattices via strains.

\begin{figure}[t]
\begin{center}
\includegraphics[width=8.5cm]{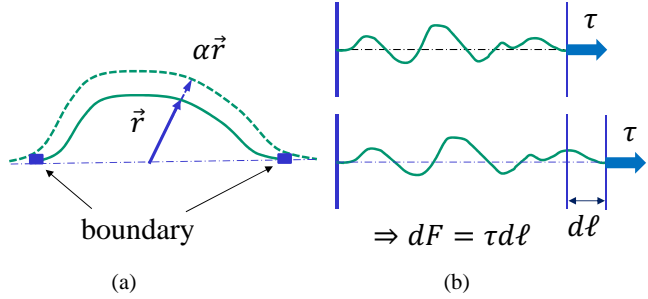}
 \caption{(Color online) (a) An illustration of a surface (denoted by ${\bf r}$) spanning a fixed boundary and its scaled surface (denoted by $\alpha{\bf r}$), where the boundary frame remains fixed,  (b) illustrations for the energy increment ($dF\!=\!\tau d\ell$) accumulated in the surface by external tensile force $\tau$. In this surface extension, the increment of projected area is given by $dA_p=d\ell$, where the side length is fixed to $1$ and remains unchanged for simplicity. }
 \label{fig-2}
\end{center}
\end{figure}
The integration variable ${\bf r}$ in $Z$ of Eq. (\ref{part-func}) can be changed to  ${\bf r}\to \alpha{\bf r}$, where $\alpha$ is called the scale parameter, and we have the scaled partition function $Z[\alpha;A_p(\alpha)]$ such that \cite{WHEATER-JP1994}
\begin{eqnarray}
\label{scaled-part-func}
\begin{split}
Z[\alpha;A_p(\alpha)]=&\alpha^{3(N-N_B)+2N_B}\\
&\sum_\sigma \int \prod_{i=1}^{N-N_B} d{\bf r}_i \int \prod_{i=1}^{N_B} d{\bf r}_i C(A_p(\alpha))\exp\left[ -\beta S(\alpha{\bf r},\sigma) \right],  \\
&S(\alpha{\bf r},\sigma)=\alpha^2 S_1+\kappa S_2+ \lambda S_F + \delta S_{\rm DM} + S_B.
\end{split}
\end{eqnarray} 
The second argument $A_p(\alpha)$ in $Z[\alpha;A_p(\alpha)]$ is given by $A_p(\alpha)\!=\!\alpha^{-2} A_p$, which comes from the assumption that the projected area $A_p$ remains fixed while the scale of the lattice is changed by $\alpha$ (Fig. \ref{fig-2}(a)). 
The factor $\alpha^{3(N-N_B)+2N_B}$ in the right hand side of Eq. (\ref{scaled-part-func}) comes from the integration measures, and only $S_1$ changes to $\alpha S_1$ in $S$ because all other terms in $S$ are scale independent. Since $Z$ is independent of $\alpha$, we have $Z[\alpha; A_p(\alpha)]=Z[1;A_p(1)]$. From this, we obtain $d Z[\alpha;A_p(\alpha)]/d\alpha|_{\alpha=1}\!=\!\left[\partial Z/\partial \alpha + (\partial Z/\partial A_p)(d A_p/d \alpha)\right]_{\alpha=1}\!=\!0$. Dividing both sides of this equation by $Z$, we have 
\begin{eqnarray}
\label{calculation-1}
3N-N_B - 2\left[\sum_\sigma \int \prod_i d{\bf r}_i C(A_p)\beta S_1 \exp(-\beta S)+A_p \frac {\partial Z}{\partial A_p}  \right] Z^{-1} =0.
\end{eqnarray} 
The expression $\partial Z/\partial A_p$ corresponds to the derivative of the constraint $C(A_p)$. 
To calculate $\partial Z/\partial A_p$, we assume for the surface with the projected area $A_P$ that the free energy $F$ is given by 
\begin{eqnarray}
\label{Z-for-surface}
F(A_p)=\tau \int_{A_0}^{A_p} dA,
\end{eqnarray} 
where $\tau$ is called frame tension because the surface spans the fixed boundary frame with projected area $A_p$. 
We should note that the expression in Eq. (\ref{Z-for-surface}) for the free energy can be called macroscopic energy. The other macroscopic energies such as bending energy corresponding to $S_2$ in Eq. (\ref{model}) are not included in this free energy, because such  bending energy is not explicitly dependent on $A_p$. To the contrary, the Hamiltonian $S$ in Eq. (\ref{model}) is a microscopic energy, which is defined by the sum of local interaction energy between the neighboring vertices or triangles $i$ and $j$. In this microscopic formulation, the function $C(A_p)$ is a constraint on the multiple integration and too complex to be expressed with $A_p$ as mentioned above.  However, this constraint on $A_p$ turns to be very simple in the macroscopic energy in Eq. (\ref{Z-for-surface}).  
For this reason, the free energy in Eq. (\ref{Z-for-surface}) is used to calculate the frame tension $\tau$ as a macroscopic physical quantity from the microscopic perspective \cite{WHEATER-JP1994}. 
Thus, by using $Z\!=\!\exp (-\beta F)$, we have $\partial Z/\partial A_p\!=\!-\beta\tau Z$. Using the symbol $\langle S_1\rangle$ for the first term of $[\cdots]$ in the left hand side of Eq. (\ref{calculation-1}), we obtain  $\tau\!=\![{2\beta\langle S_1\rangle\!-\!(3N\!-\!N_B)}]/({2\beta A_p})$, and hence we have the formula for the frame tension $\tau$ such that  \cite{WHEATER-JP1994}
\begin{eqnarray}
\label{frame-tension}
\tau=\frac{\langle S_1\rangle-(3N-N_B)T}{2A_p}. 
\end{eqnarray} 
We should note that $\langle S_1\rangle$ implicitly depends on $T$. Note also that this $\tau$ can be written as $\tau\!=\![\langle S_1\rangle\!-\!\langle S_1^0\rangle]/A_p$, where $\langle S_1^0\rangle$ is $\langle S_1\rangle$ for $\tau\!=\!0$. In the calculation of $\tau$, we use the formula in Eq. (\ref{frame-tension}), in which $\langle S_1^0\rangle$ is not necessary.

Note that the projected area $A_p$ used in $\tau$ is different from the real projected area of the lattice, though the difference of $A_p$ and the real projected area of the lattice is expected to be very small. The projected area $A_p$ of the boundary frame is the only area that is connected with experimental measurements of $\tau$. This comes from the fact that the elastic energy by tensile force is accumulated in the surface is given by "force $\times$ increment of the projected area" (Fig. \ref{fig-2}(b)). 
The important point to note is that the {$A_p$} in Eq. {(\ref{Z-for-surface})} simply the area of boundary frame.

\subsection{Monte Carlo technique\label{MC-technique}}
The spins $\sigma$ and the vertex positions ${\bf r}$ are updated using the standard Metropolis MC technique \cite{Metropolis-JCP-1953,Landau-PRB1976}.  A new $\sigma_i^\prime$ at the vertex $i$, generated independently of the old $\sigma_i$ by using three different uniform random numbers, is accepted with the probability ${\rm Min}[1, \exp -{\it \Delta}S]$, where ${\it \Delta}S\!=\!S({\rm new})\!-\!S({\rm old})$. We should note that the three different uniform random numbers $\sigma_{x,y,z} \in (-0.5,0.5]$ are generated with the constraint $\sigma_x^2 + \sigma_y^2 + \sigma_z^2\leq 0.25$. This constraint makes the point $(\sigma_x,\sigma_y,\sigma_z)$ uniform in the ball of radius $0.5$.  Then, the vector $(\sigma_x,\sigma_y,\sigma_z)$ is normalized such that $\sigma_x^2 + \sigma_y^2 + \sigma_z^2 = 1$. Thus, the distribution of this unit vector $(\sigma_x,\sigma_y,\sigma_z)$ is expected to be uniform on the unit sphere. Because of this independence of $\sigma_i^\prime$ on the old $\sigma_i$, the rate of acceptance is not controllable. A new ${\bf r}_i^\prime\!=\!{\bf r}_i\!+\!{\it \Delta}{\bf r}$ is also updated in the probability ${\rm Min}[1, \exp -{\it \Delta}S]$. In this ${\bf r}_i^\prime$, the symbol ${\it \Delta}{\bf r}$ denotes a random three dimensional vector inside a small sphere of radius $R_0$, which is determined such that the rate of acceptance for ${\bf r}_i^\prime$ is approximately $50\%$. One Monte Carlo sweep (MCS) consists of a simultaneous updating ${\bf r}$ and ${\sigma }$ at a given spin and repeating this for all spins.

These MC updates for $\sigma$ and ${\bf r}$ are started after a generation of skyrmion configuration, which is the ground state (GS) obtained by the technique described in \cite{Hog-etal-JMagMat2018}.  This GS is generated by the iteration for $\sigma$, and during this iteration, the variables ${\bf r}$ are fixed to the initial configuration, like in the snapshot in Fig. \ref{fig-1}(a). Therefore, the GS  is influenced by both of thermal fluctuations of spins and surface fluctuations after the start of MC updates for $\sigma$ and ${\bf r}$. However, since the GS is determined depending on the parameters $\lambda$, $\delta$ and $B$, it is close to the equilibrium state for sufficiently low  $T$ and sufficiently large $\kappa$. 
The MC simulation starting with the GS configuration is called "slow heating of the system from the GS''.

This procedure is used in strongly degenerate systems, strongly disordered systems, systems with many meta-stable states and systems with competing interactions if we know the GS. Starting with initial random configurations, namely cooling the system from high $T$, except for simple systems with long-range ordering and small unit cell such as ferromagnets, it is well known that MC simulations encounter problems to equilibrate the system at low temperatures. This is not a problem of simulation time. This comes from the fact that the system is often stuck in meta-stable states at low $T$. One way to get rid of this is to determine the GS by other methods such as the steepest-descent method used in Ref. \cite{Hog-etal-JMagMat2018} and slowly heat the system from its GS.   Even in this case, the relaxation time is very long. This happens in strongly non-uniform systems such as spin glasses or systems with huge unit cells such as skyrmion crystals  \cite{Hog-etal-JMagMat2018} .  

In general, when the GS has several degenerate configurations, only one of them is chosen by the system if we are able to cool the system down to $T\!=\!0$. But this is impossible for complex systems as said above. What we do is to choose a ground state and make a slow heating. To see if the system is equilibrated or not, we have to make a finite-time scaling to deduce properties at the infinite time. This is very similar in spirit with the finite-size scaling used to deduce properties at the infinite crystal size. We have previously performed a finite-time scaling for the 2D skyrmion crystal of the rigid lattice \cite{Hog-etal-JMagMat2018}. We have seen that skyrmions need much more than $10^6$ MC steps per spin to relax to equilibrium.  The order parameter follows a stretched exponential law as in spin glasses and stabilized at non-zero values for $T\!<\!T_c$ at the infinite time.
If there is no skyrmion stability, we would not have non-zero values of the order parameter for  $T\!<\!T_c$ at the infinite time. 
Based on the work of Ref. \cite{Hog-etal-JMagMat2018} on the same model without elasticity, we have taken sufficient run times as indicated below.
We note that in the present work, as soon as $T$ is not zero, the non-uniform deformation of the lattice excludes invariance by global rotation.

The lattice size for the simulations is $N\!=\!10^4 (\Leftrightarrow L\!=\!100)$. This lattice size is relatively small compared to those (lattices up to $L\!=\!800$) in Ref. \cite{Hog-etal-JMagMat2018}, where the phase transition between the skyrmion and paramagnetic phases is studied. However, we do not go into details in the phase transitions in this paper, and therefore, the lattice size $N\!=\!10^4$ is considered to be sufficient. The total number of MCSs is $3\times 10^8\sim 2.9\times 10^9$ including $1\times 10^7\sim 2.3\times 10^9$ thermalization MCSs, and physical quantities are calculated every 1000 MCSs.  This total number of MCSs depends on the fluctuations of lattice, and it is small (large) for simulations on the lattices with small (large) fluctuations. The reason for this dependence is that the skyrmion configuration is almost uninfluenced (considerably influenced) if the lattice fluctuations are relatively small (large).

\section{Results under isotropic strain\label{results-isotropic}}
\subsection{Phase diagram and Snapshots\label{phase-diagram-snapshots-iso}}
\begin{figure}[t]
\begin{center}
\includegraphics[width=11.5cm]{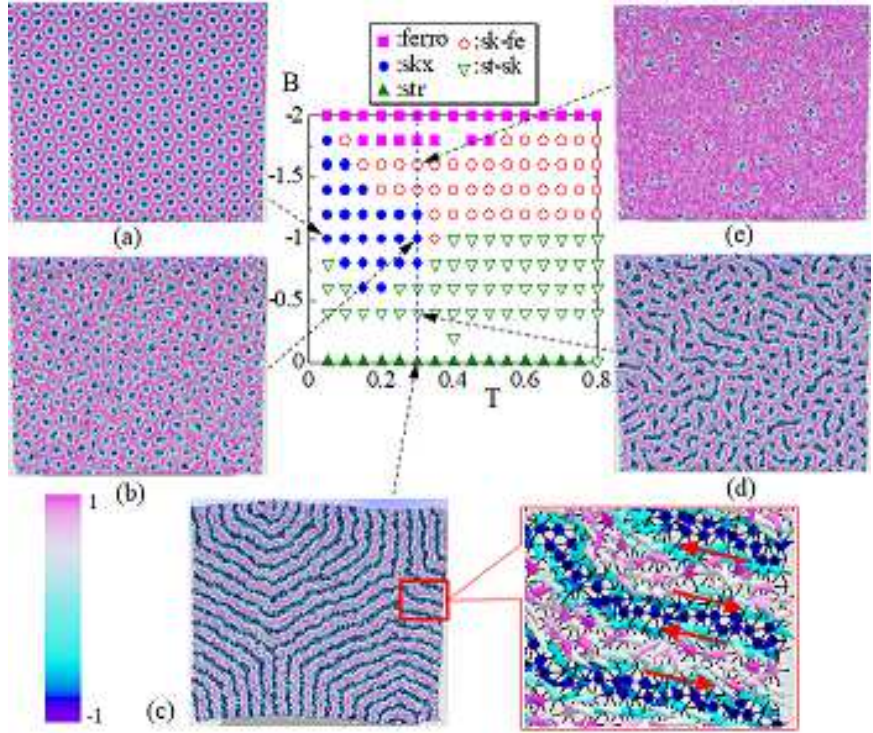}
 \caption{(Color online) A phase diagram on the surface $B$ vs. $T$, with snapshots at (a) skx, (b) skx, (c) str, (d) st-sk, and (e) sk-fe phases, and the color variation corresponding to $\sigma_z$.  The results are obtained on the lattice of $a\!=\!1$ and with the parameters $\kappa\!=\!25$, $\lambda\!=\!0.5$ and  $\delta\!=\!1$. The vertical dashed line is drawn at $T\!=\!0.3$, where the skx phase terminates when $T$ is increased. The variable $\sigma$ is drawn by a cone (cylinder) in the enlarged (original) snapshot. The arrows in the enlarged snapshot denote the direction of $\sigma$. The lattice size is $N\!=\!10^4$. }
 \label{fig-3}
\end{center}
\end{figure}
Our focus is on the frame tension, which is expected to be influenced by spin configurations because of the DM interaction as mentioned in Section \ref{Sky-model}. To study this, we firstly show a phase diagram on the $B\!-\!T$ plane in Fig. \ref{fig-3}, where snapshots obtained at several points are also shown. The parameters $a$, $\lambda$ and $\kappa$ are fixed to $a\!=\!1$, $\lambda\!=\!0.5$, and $\kappa\!=\!25$. In the figure, the letters ferro, skx and str denote the FM, skyrmion crystal and stripe phases, and sk-fe and st-sk denote the intermediate phases between these three different phases. The stripe pattern is identical with that observed in \cite{Hog-etal-JMagMat2018,Yu-etal-Nature2010,Yu-etal-PNAS2012,Kwon-etal-JMagMat2012}. The typical configuration of skx is shown in Fig. \ref{fig-3}(a) at $T\!=\!0.05$, which is sufficiently low. The criteria for distinguishing one from another is as follows: Along the vertical dashed line at $T\!=\!0.3$, we start from the skx phase at $B\!=-1$ (Fig. \ref{fig-3}(b)) and go downwards, then skyrmions start to connect and form oblong islands. This phase is the st-sk phase denoted by $\bigtriangledown$. The length of islands increases when $|B|$ decreases, and for the sufficiently small $|B|$, the islands are connected to each other like in Fig. \ref{fig-3}(c), which we call the stripe phase. In the enlarged snapshot of Fig. \ref{fig-3}(c), we see that the direction of spins along the stripes changes alternatively \cite{Yu-etal-Nature2010,Yu-etal-PNAS2012,Kwon-etal-JMagMat2012}. The configuration typical to the st-sk is shown in Fig. \ref{fig-3}(d) at $B\!=\!-0.4$.  To the contrary, if we start from the skx phase at $B\!=\!-1$ and go upwards along the dashed line, we find that the total number of skyrmions decreases, and this is the sk-fe phase like in the snapshot of Fig. \ref{fig-3}(e), where $B\!=\!-1.6$. For sufficiently large $|B|$, the skyrmion configuration completely disappears, and the phase changes from sk-fe to ferro denoted by the solid square symbol. 

The main purpose of this phase diagram is to find where the skyrmion appears, and for this reason, the diagram includes empty regions without symbols where no simulation is performed because skyrmions are not expected. The skyrmion phase appears for the intermediate region of $B$ and for the low temperature region as expected. At low temperatures, the lattice fluctuations are suppressed, and the skx configuration is actually smooth  at $T\!=\!0.05$ (Fig.\ref{fig-3}(a)), while at $T\!=\!0.3$, which is the highest temperature at which the skx can be seen, and the skx configuration is relatively fluctuated (Fig.\ref{fig-3}(b)). For $T\!=\!0.4$, no skx configuration appears. In the regions of $T(>0.3)$ close to $T\!=\!0.3$, the initial ground states are still skx, however, these skx states disappear after a sufficiently large number of MC sweeps for the updates of both $\sigma$ and ${\bf r}$. This is due to the thermal fluctuations of spins, which are originally expected for such a relatively large $T$ region. The surface fluctuations are not always the reason for the violation of skx configurations, because,  at the same temperature  $T\!=\!0.3$ on the rigid lattice, the skx phase is separated from the sk-fe phase (using the terminology in this paper) \cite{Hog-etal-JMagMat2018}.  We should note that the phase diagram also indicates the stability of skyrmions at a sufficiently low temperature, however, this is independent of whether the lattice is fluctuated or not, because skyrmions in the same 2D model are stable in the region $T\!\leq\! 0.3$ on the rigid lattice \cite{Hog-etal-JMagMat2018}. As mentioned above, different kinds of interactions can give rise to the birth of skyrmions. Skyrmions in our model come from the competition of only two interactions, $S_{DM}$ and $S_F$, in the ground state. That is the reason why our skyrmions appear in such low temperature region and there is only the skyrmion phase below the paramagnetic phase. In other models and in experiments, various kinds of interactions (more than two) may compete and can give rise to various phases with increasing $T$, the skyrmion phase is one of these phases and can be found at high temperature.

\begin{figure}[t]
\begin{center}
\includegraphics[width=11.5cm]{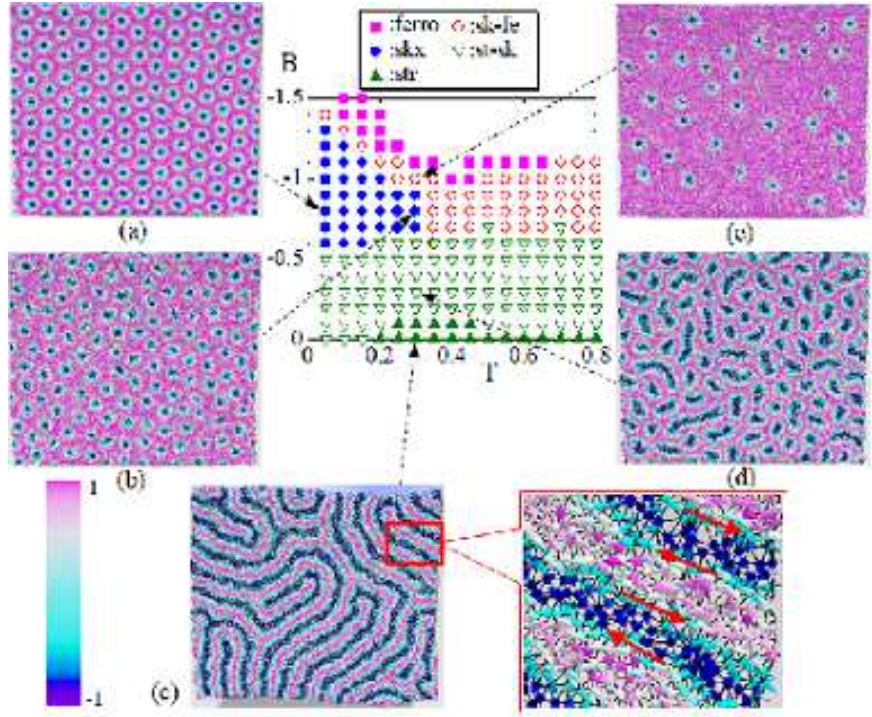}
 \caption{(Color online) A phase diagram on the surface $B$ vs. $T$, with snapshots at (a) skx, (b) skx, (c) str, (d) st-sk, and (e) sk-fe phases, and the color variation corresponding to $\sigma_z$. The results are obtained on the lattice of $a\!=\!0.75$ and with the parameters $\kappa\!=\!15$, $\lambda\!=\!1$ and  $\delta\!=\!1$. The vertical dashed line is drawn at $T\!=\!0.3$, where the skx phase terminates when $T$ is increased.  The lattice size is $N\!=\!10^4$.}
 \label{fig-4}
\end{center}
\end{figure}
The frame tension $\tau$  is obtained by fixing the projected area $A_p$ of the boundary as in Eq. (\ref{frame-tension}), and $A_p$ is  determined by the lattice spacing $a$ when the total number of vertices is fixed. Therefore, it is interesting to see the dependence of $\tau$ on $a$, and, for this purpose, we change the lattice spacing from $a\!=\!1$ to $a\!=\!0.75$,  and, as a result, the side length of the lattice is also reduced by $3/4$ from the lattice used in Fig. \ref{fig-3}. It is also interesting to see whether the transition temperature $T\!=\!0.3$, between the skx and sk-fe phases in Fig. \ref{fig-3}, changes depending on the lattice spacing $a$. To further increase the surface fluctuations, we also reduce the bending rigidity to $\kappa\!=\!15$ from $\kappa\!=\!25$ assumed for Fig. \ref{fig-3}.

We find from Fig. \ref{fig-4} that the phase structure is almost the same as that in Fig. \ref{fig-3}. The temperature separating the skx and sk-fe phases remains the same as $T\!=\!0.3$. This implies that the surface fluctuation, if it is small enough, does not as strongly influence the phase structure of skyrmions.

\subsection{Frame tension and interaction energies\label{frame-tension-iso}}
\begin{figure}[t]
\begin{center}
\includegraphics[width=9.5cm]{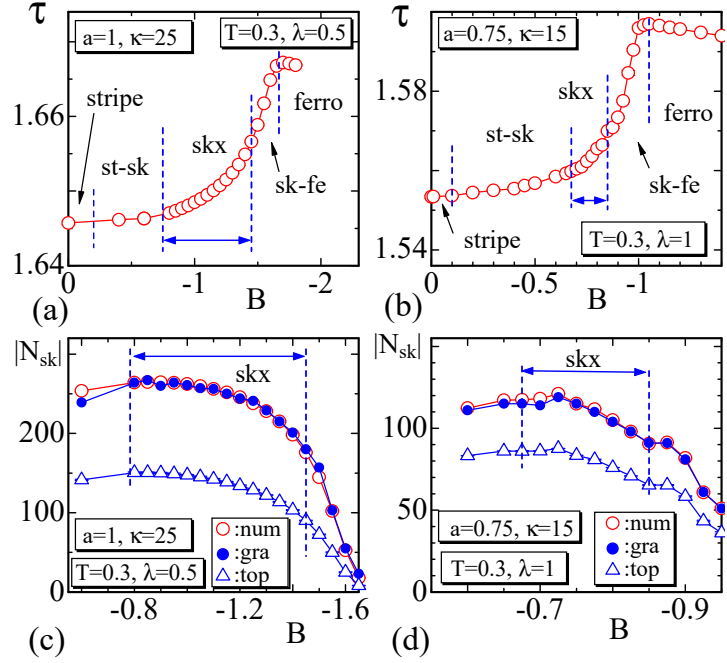}
 \caption{(Color online) (a),(b) The frame tension $\tau$ vs. $B$ obtained along the dashed line on the phase diagram in Figs. \ref{fig-3} and \ref{fig-4}, (c),(d) the total number of skyrmions $|N_{\rm sk}|$ in the skx phase obtained along the same dashed lines, where $|N_{\rm sk}|$ is obtained by the three different techniques "num", gra" and "top". The numerical technique "num" denoted by the symbol ($\bigcirc$) is not always correct outside the skx phase. Inside the skx phase in both (c) and (d), $|N_{\rm sk}|$  increases with decreasing $\tau$ ($\Leftrightarrow$ increasing $p$) except the boundary region close to the st-sk in (d). }
 \label{fig-5}
\end{center}
\end{figure}
Now we plot the frame tension $\tau$ vs. the magnetic field $B$ in Figs. \ref{fig-5}(a) and \ref{fig-5}(b). The data are obtained along the dashed lines on the phase diagrams in Figs. \ref{fig-3} and \ref{fig-4}. The vertical dashed lines on Figs. \ref{fig-5}(a) and \ref{fig-5}(b) represent the phase boundaries. As we find from the figures, the value of $\tau$ in Fig. \ref{fig-5}(a) is larger than that in Fig. \ref{fig-5}(b). This is because, as explained in the previous subsection, the projected area of the lattice in Fig. \ref{fig-5}(a) is larger than that in Fig. \ref{fig-5}(b), and consequently the surface in Fig. \ref{fig-5}(a) is relatively expanded and smooth compared with that in Fig. \ref{fig-5}(b). We also find in both cases that $\tau$ increases with increasing $|B|$ or equivalently $\tau$ decreases with decreasing $|B|$ in the skx phase and also in the sk-fe phase.

The total number $|N_{\rm sk}|$ of skyrmions plotted in Figs. \ref{fig-5}(c) and \ref{fig-5}(d) is calculated/measured in three different techniques. The first one is to calculate the topological skyrmion number
\begin{eqnarray}
\label{sky_number}
N_{\rm sk}=\frac{1}{4\pi} \int d^2x \sigma \cdot \frac{\partial \sigma}{\partial x_1}\times \frac{\partial \sigma}{\partial x_2}
\end{eqnarray}
numerically on triangulated surfaces. The plus or minus sign of $N_{\rm sk}$ depends on that of $B_z$, and hence, we use the absolute symbol $|\;\;|$ for the total number of skyrmions. The discrete form of $N_{\rm sk}$ in Eq. (\ref{sky_number}) is expressed by
\begin{eqnarray}
\label{sky_number_disc}
N_{\rm sk}=\frac{1}{4\pi} \sum_{\it \Delta} (\sigma_2-\sigma_1)\times(\sigma_3-\sigma_1),
\end{eqnarray}
where $\sigma_i$ denotes $\sigma$ at the vertex $i$ of the triangle ${\it \Delta}$ (Fig. \ref{fig-6}(a)). In this discretization, the integration $\int d^2x$ and derivatives $\partial_i \sigma$ are simply replaced by $\int d^2x\to\sum_{\it \Delta}$ and  $\partial_1 \sigma\to \sigma_2\!-\!\sigma_1$,  $\partial_2 \sigma\to \sigma_3\!-\!\sigma_1$. We should note that there are three possible local-coordinate origins in one triangle, and hence summing over the contributions to $N_{\rm sk}$ from these three coordinates with the numerical factor $1/3$, we calculate $N_{\rm sk}$ in Eq. (\ref{sky_number_disc}). 
The obtained value is denoted by "top" on Figs. \ref{fig-5}(c) and \ref{fig-5}(d). No numerical cutoff is necessary for this calculation of $N_{\rm sk}$ by Eq. (\ref{sky_number_disc}).  
The second technique to calculate $N_{\rm sk}$ is as follows: In the case $B_z<0$, we assume that there uniquely exists a vertex $i$ such that $\sigma^z_i<\sigma^z_j$  for all vertices $j(\not=\!i)$ inside the range up to three-nearest neighbors in the skx phase (Fig. \ref{fig-6}(b)). Here, the condition $\sigma^z_i<0$ is imposed on the vertex $i$.  This assumption for the uniqueness of $i$ for identifying the minimum $\sigma_z$ is reasonable, because the spins in the skyrmion configuration have values uniformly on the unit sphere. Thus, we expect that the vertices determined in this technique are in one-to-one correspondence with skyrmions. Therefore, $|N_{\rm sk}|$ is obtained by counting these uniquely-determined vertices. The obtained vcalue of $|N_{sk}|$ by this numerical technique is denoted by "num" on the figures.  The third technique to obtain $|N_{\rm sk}|$ is a graphical technique, which uses a graphics software. The detailed information of this graphical technique is written in Appendix \ref{appendix-A}. The corresponding data  plotted on Figs. \ref{fig-5}(c) and \ref{fig-5}(d) are denoted by "gra". 

\begin{figure}[t]
\begin{center}
\includegraphics[width=8.5cm]{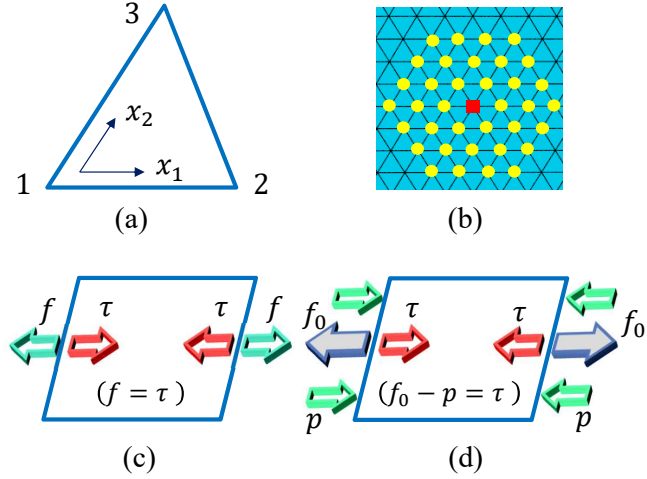}
 \caption{(Color online) (a) A triangle $123$ with local coordinate axes $x_1$ and $x_2$, (b)  A vertex (solid square) and its three-nearest neighbor vertices (solid circles) for counting the total number of skyrmions.  (c) External force $f$ and material response $\tau$ are balanced, where $f$ is increased (decreased) then $\tau$ is increased (decreased) and vice versa.  (d) $f$ can be replaced by $f_0-p$ under constant $f_0$ using a pressure $p$, where $p$ is increased (decreased) then $\tau$ is decreased (increased) and vice versa.  Units of these forces $\tau$, $f$ and $p$ are assumed to be the same. }
 \label{fig-6}
\end{center}
\end{figure}

The results obtained by the techniques "num" and "gra" are almost the same at least in the skx and sk-fe phases. Moreover, from Fig. \ref{fig-5}(c), we find that the $|N_{sk}|$ obtained by "num" and "gra" decrease to $|N_{sk}|\!\to\!0$ for sufficiently large $|B_z|$ close to the ferro phase. This implies that "num" and "gra" correctly count the skyrmions except the stripe phase, where  $|N_{sk}|$ of "num" slightly different from that of "gra" (see Fig. \ref{fig-5}(c)). Indeed,  $|N_{sk}|$ obtained by both "gra" and "num" are unreliable in the st-sk phase because a long island is counted as skyrmions by these techniques. On the other hand, the results obtained by the technique "top" are smaller than those by the other two techniques. This error observed  in the technique "top" is simply due to the numerical errors caused by the discretization technique described above, though this discretization is simple enough and straightforward.

We consider that the results by "top" are reliable not only in the skx phase but also in the other phases except the magnitude of $|N_{sk}|$. Thus, apart from the precise value of  $|N_{sk}|$, the variation of $|N_{sk}|$ vs. $B$ in the skx phase is correctly obtained, because the results of the three techniques on this point are consistent to each other.  This consistency on the the variation of  $|N_{sk}|$ vs. $B$ is sufficient for our purpose in this paper.

To compare the results of $\tau$ with the experimental data in \cite{Nii-etal-NatCom2015}, we should replace $\tau$ by the corresponding external pressure $p$, because the experimental data are expressed by using pressure. We should remind ourselves of the fact that $\tau$ is a response of the material to the applied external pressure $p$.  Therefore, the replacement of $\tau$ with $p$ is possible using the action-reaction principle or the equilibrium principle of forces. Recalling that $\tau$ is balanced with an external mechanical force $f$ applied to the boundary in this case, we understand that the positive $\tau$ implies that the corresponding external force $f$ is a tensile force (Fig. \ref{fig-6}(b)). Units of $f$ and $\tau$ are assumed to be the same (this means that $\tau L$ is written as $\tau$ for simplicity). Moreover, a decrease of the tensile force $f$ can be understood as a result of the fact that a small pressure $p(>\!0)$ is applied in addition to the constant tensile force $f_0(>\!0)$ such that $f=f_0-p$, where the unit of $p$ is also assumed to be same as that of $f$ (Fig. \ref{fig-6}(c)). Thus, the increment (decrement) of $\tau$ is equivalent with the decrement (increment) of $p$. To summarize, the frame tension $\tau$ is identified with the external tensile force $f$, and this $f$ can be replaced by $-p$ using the constant $f_0$ such that $f=f_0\!-\!p$.  

\begin{figure}[t]
\begin{center}
\includegraphics[width=9.5cm]{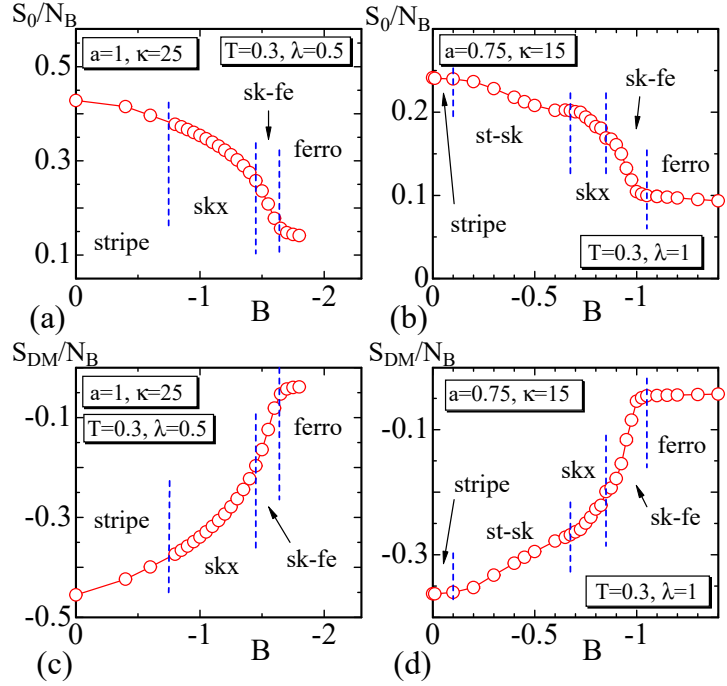}
 \caption{(Color online) (a),(b) The FM energy $S_0/N_B$ vs. $B$ obtained along the dashed line on the phase diagram in Figs. \ref{fig-3} and \ref{fig-4}, (c),(d) the DM interaction energy $S_{\rm DM}$ along the same dashed lines.    }
 \label{fig-7}
\end{center}
\end{figure}
Using this $p$, the observed fact that the decrease of $\tau$ in the skx phase as $|B|$ decreases (from the right to left on the $B$ axis) in Figs. \ref{fig-5}(a),(b) should be understood such that an increase of $p$ is caused by a decrease of $|B|$. At the same time, we find from Figs.  \ref{fig-5}(c),(d) that $N_{\rm sk}$ increases with decreasing $|B|$ in the relatively large $|B|$ region in the skx phase indicated by the arrows ($\leftrightarrow $). In this sense, it is also understood that the sk-fe phase is changed to the skx phase by an increment of $p$, which is caused by a decrease of  $|B|$.  
To summarize, our results show that  $N_{\rm sk}$ increases with increasing $p$ ($\leftrightarrow $ decreasing $\tau$) and a topological phase transition from the sk-fe to skx also occurs when $p$ is increased. 
 Thus, we find that this numerically-observed phenomenon is qualitatively consistent with the experimentally-observed fact that skyrmions are created by a lateral pressure to a 3D material \cite{Nii-etal-NatCom2015}.

The FM energy $S_0/N_B$ vs. $B$ and the DM interaction $S_{\rm DM}$ vs. $B$ are plotted in Figs.  \ref{fig-7}(a),(b), and  \ref{fig-7}(c),(d), where $N_B$ is the total number of bonds. We find that both energies rapidly change  in the sk-fe phase or at the boundary between the skx and sk-fe phases.

\begin{figure}[t]
\begin{center}
\includegraphics[width=9.5cm]{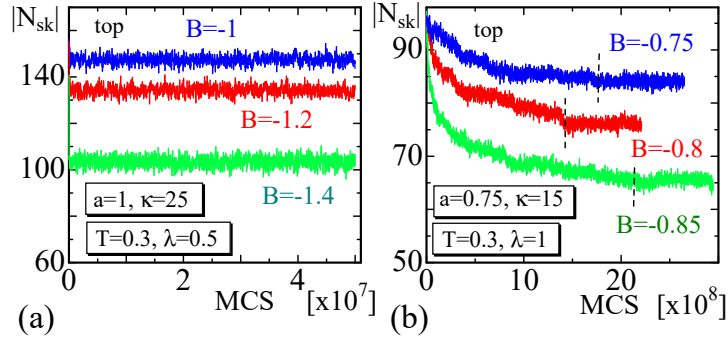}
\caption{(Color online) The total skyrmion number $|N_{sk}|$ vs. MCS for (a)  $a\!=\!1$ and $\kappa\!=\!25$ and (b)  $a\!=\!0.75$ and $\kappa\!=\!15$, where  $|N_{sk}|$ is the discrete topological charge calculated by Eq.(\ref{sky_number_disc}).  The convergence speed of  $|N_{sk}|$ in (b) is relatively low compared to that in (a) and depends on the magnetic field $B$, where $B\!=\!-0.85$ is close to the phase boundaries between the skx and sk-fe phases. \label{fig-8}}
\end{center}
\end{figure}
Next, the convergence speed of simulations is discussed. In Fig. \ref{fig-8}(a), we plot $|N_{sk}|$ vs. MCS obtained at  $a\!=\!1$ and $\kappa\!=\!25$, where $|N_{sk}|$ denoted by "top" on the figures indicates that $|N_{sk}|$ is calculated by the discrete topological charge in Eq. (\ref{sky_number_disc}). Note that the $|N_{sk}|$ calculated by "num" shows the same behavior in their variations as the one plotted in Figs. \ref{fig-8}(a),(b), though the values $|N_{sk}|$ themselves are different from each other as demonstrated in Figs.  \ref{fig-5}(c),(d). Therefore, we use $|N_{sk}|$ by "top" vs. MCS to find the thermalization MCS. From these plots in Figs. \ref{fig-8}(a),(b),  we find that the skyrmion configurations remain almost unchanged during MC updates at least in the case of $a\!=\!1$ and $\kappa\!=\!25$, where the initial configurations immediately change to those equilibrium configurations in the early stage of MC iterations. In contrast, we find that the convergence is very slow for $a\!=\!0.75$ and $\kappa\!=\!15$ from the plots in Fig. \ref{fig-8}(b).  In this case, the surface fluctuation is relatively large and destroys the skyrmion configurations, however, the skx phase becomes stable after sufficiently large number of MCSs. This implies that the skx phase is stable even on surfaces where the surface fluctuation is considerably large compared to the case  $a\!=\!1$ and $\kappa\!=\!25$. The total number of MCS for thermalization is $1\times 10^7$, which is fixed independent of $B_z$, for $a\!=\!1$ and $\kappa\!=\!25$, while it is determined depending on $B_z$ for $a\!=\!0.75$ and $\kappa\!=\!15$. The vertical dashed lines denote the thermalization MCSs in Fig. \ref{fig-8}(b). Since the configuration for $B\!=\!-0.85$ is close to the phase boundary between the skx and sk-fe  phases, the convergent speed is relatively slow compared to the other two in Fig. \ref{fig-8}(b).

\section{Results under uniaxial strains\label{results-anisotropic}}
\begin{figure}[t]
\begin{center}
\includegraphics[width=8.5cm]{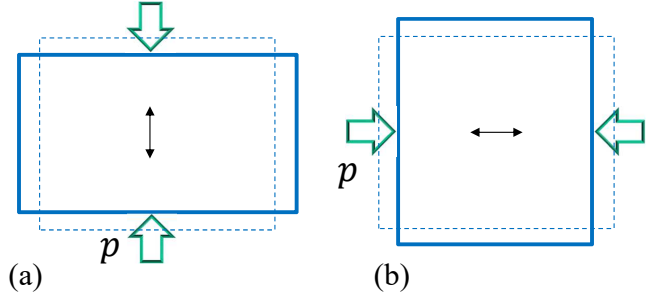}
 \caption{(Color online) A boundary shape corresponding to the constraints for (a) $\xi<1$ and (b) $\xi>1$ in Eq. (\ref{latt-space-xy}).  The shapes of boundary in (a), (b) correspond to the conditions that additional stresses $p$ are applied along $L_2$  ($\updownarrow $)  and $L_1$ ($\leftrightarrow $)  directions, respectively. The "strain relief" direction is same as the $p$ direction in both (a) and (b).}
 \label{fig-9}
\end{center}
\end{figure}
It is also interesting to see the results obtained under uniaxial stresses, although the 2D simulation results cannot always be compared to those of experiments where 3D materials are always targeted. To realize the uniaxial stress condition, we deform the edge length of the boundary by changing the lattice spacing $a$ to be direction dependent such that $(a_1,a_2)\!=\!(a/\xi,a \xi)$ as mentioned in the Introduction. We should note that this constraint can be called "uniaxial strains".  Indeed, by this modification of lattice spacing from $a$ to $a_1$ ($a_2$) in $L_1$ ($L_2$) direction in Eq. (\ref{latt-size}), we effectively have the modified edge length in the unit of $a$ such that
\begin{eqnarray}
\label{latt-space-xy}
(\xi^{-1}L_1,\xi L_2).
\end{eqnarray}
We should note that the projected area remains unchanged under this modification.  It is more straightforward to use these effective edge lengths $(\xi^{-1}L_1,\xi L_2)$ for the direction dependent lattice spacing $(a_1,a_2)$, because the lattice deformation by uniaxial strains is simply expressed by $\xi$ in this expression. Note also that both $L_1$ and $L_2$ originally denote the total number of lattice points, and the expression in Eq. (\ref{latt-space-xy}) has the meaning of edge lengths only when $a$ is multiplied.

The condition $\xi\!<\!1$ corresponds to the case that the edge length $\xi^{-1} L_1$ ($\xi L_2$) is longer (shorter) than its original $L_1$ ($L_2$), and the corresponding tensile force along  $L_1$ ($L_2$)  direction is expected to be smaller (larger) than that along $L_2$ ($L_1$) direction. Therefore, by using the pressure $p$ in Fig. \ref{fig-6}(d), we understand that $\xi\!<\!1$  ($\xi\!>\!1$) corresponds to the condition that a uniaxial stress $p$ is applied along $L_2$ ($L_1$) direction as shown in Figs. \ref{fig-9}(a) and \ref{fig-9}(b). 

Note also that the "strain relief" direction is identical with the $p$ direction in both Figs. \ref{fig-9}(a) and \ref{fig-9}(b). The reason is as follows:  The surface is expanded by a strong external force $f$ at the boundary frame in both $L_1$ and $L_2$ directions as described in Section \ref{frame-tension} (see also Fig. \ref{fig-6}(c)). Therefore, if an additional pressure $p$ is applied to the $L_2$ direction, the side length along the $L_2$ direction is slightly reduced like in Fig. \ref{fig-9}(a), and, as a consequence, the corresponding stress $\tau$ is also "relieved" such that $\tau=f\!-\!p$.

\begin{figure}[t]
\begin{center}
\includegraphics[width=12.5cm]{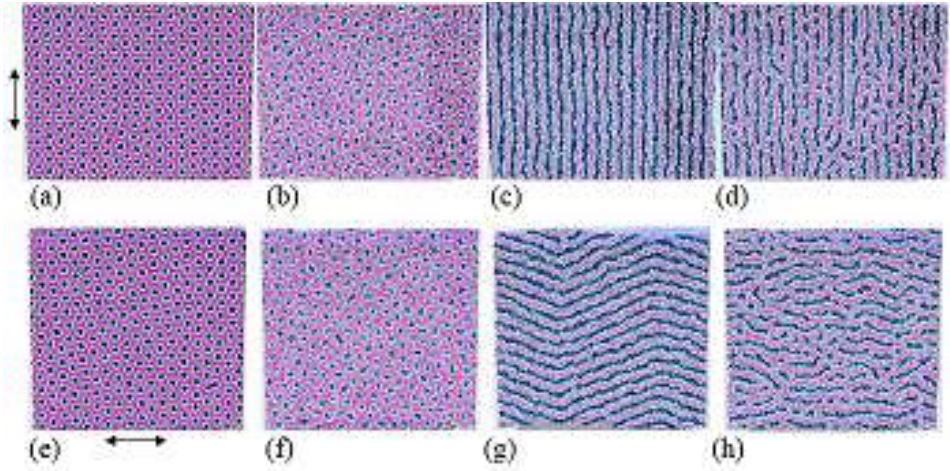}
 \caption{(Color online) Snapshots of surfaces, where an effective stress is applied to the directions ($\updownarrow $) for (a),(b),(c),(d) and ($\leftrightarrow $) for (e),(f),(g),(h). The parameter $\xi$ in Eq. (\ref{latt-space-xy}) is $\xi^2\!=\!0.9(\Leftrightarrow \xi\!\simeq\!0.9487)$ for (a)--(d)  and $\xi^2\!=\!1.1(\Leftrightarrow \xi\!\simeq\!1.049)$ for (e)--(h). Both (a)--(d) and (e)--(h) correspond to Figs. \ref{fig-3}(a)--\ref{fig-3}(d), where the lattice spacing is $a\!=\!1$ and the parameters are $\kappa\!=\!25$,  $\lambda\!=\!0.5$ and  $\delta\!=\!1$.  }
 \label{fig-10}
\end{center}
\end{figure}
In Fig. \ref{fig-10}, we show snapshots obtained under $\kappa\!=\!25$, $T\!=\!0.3$, $\lambda\!=\!0.5$, which are the same as those for Fig. \ref{fig-3}. The parameter $\xi$ is fixed to $\xi^2\!=\!0.9(\Leftrightarrow \xi\!\simeq\!0.9487)$ for Figs. \ref{fig-10}(a)--\ref{fig-10}(d) (upper row) and $\xi^2\!=\!1.1(\Leftrightarrow \xi\!\simeq\!1.049)$ for Figs. \ref{fig-10}(e)--\ref{fig-10}(h) (lower row). The four snapshots in the upper (lower) row are obtained under $B\!=\!-1$, $B\!=\!-1$, $B\!=\!0$, $B\!=\!-0.4$  ($B\!=\!-0.8$, $B\!=\!-0.8$, $B\!=\!0$, $B\!=\!-0.3$),  which correspond to Figs.\ref{fig-3}(a)--\ref{fig-3}(d)  (Figs.\ref{fig-4}(a)--\ref{fig-4}(d)). As mentioned above, $\xi^2\!=\!0.9$  corresponds to that of an additional uniaxial stress being applied to $L_2$ ($\Leftrightarrow$ vertical ($\updownarrow $)) direction, and we find from Figs. \ref{fig-10}(a),(b) that the skyrmion phase is not influenced by this uniaxial stress, while the stripe domains including islands align along this direction. In contrast,  $\xi^2\!=\!1.1$ corresponds to that of an additional uniaxial stress being applied in the $L_1$ ($\Leftrightarrow$  horizontal ($\leftrightarrow $)) direction, and we find that the results are consistent with those in the case for $L_2$ direction. Indeed,  we see no influence on the skyrmion phase (Figs. \ref{fig-10}(a) and \ref{fig-10}(e)) except one pair of skyrmions merged in the down-right corner of Fig. \ref{fig-10}(e), the configuration of which is only slightly influenced by the strain because it is obtained at the boundary between skx and st-sk phases. The stripe domains almost align along the stress direction  (Figs. \ref{fig-10}(c),(d) and \ref{fig-10}(g),(h)). Almost the same results are obtained on rigid lattices with uniaxial strain conditions. Therefore, we find  no remarkable influence of lattice fluctuations on the skyrmion configurations.  

The stripe directions in Figs. \ref{fig-10}(c),(d) and \ref{fig-10}(g),(h) are different from
the experimental results in Ref. \cite{JDho-etal-APL2003}. One possible explanation is that skyrmions in the experimental material may have another physical origin than the DM interaction. We know that skyrmions can be generated from a DM interaction as in the present model, by the frustration in a triangular antiferromagnet, by a long-range dipole-dipole interaction, by different kinds of anisotropies. The response of the system to an external parameter depends on the physical origin generating skyrmions.  Different materials have different origins of skyrmions; our model does not certainly correspond to materials used in Ref. \cite{JDho-etal-APL2003}. The magnetoelastic coupling included in $S_{DM}$ is isotropic and not always reflected in the responses of stripe domains under the uniaxial strains, contrary to the experimental result in Ref. \cite{Shibata-etal-Natnanotech2015}, where it is reported that the anisotropic deformation of skyrmions under uniaxial strains is caused by a direction-dependent DM interaction coefficient. Again, we emphasize that different materials have different kinds of skyrmions. Different models have to be considered. For example, the present form of $S_F$ for FM interaction does not take into account the lattice deformation.

We should note also that the stripe pattern in Fig. \ref{fig-10}(g) is different from the zigzag spin spiral observed in the Fe double layer by Hsu et.al. \cite{Hue-etal-PRL2016,Finco-etal-JRB2016}. In their experiment, the influence of uniaxial strain relief on the spin spiral was studied by STM measurements, and the authors reported that the $\vec q$ vector of the zigzag spin spiral is perpendicular to the "strain relief" direction, which is parallel to the zigzag pattern. The "strain relief" direction in our 2D model is also parallel to the stripe direction as mentioned above, as was the case of the zigzag pattern in their analysis. However, in their analysis, the direction of spins is vertical to the stripe direction, while in our 2D calculation the spin direction is parallel to the stripe direction or vertical to the 2D plane. 

\begin{figure}[t]
\begin{center}
\includegraphics[width=12.5cm]{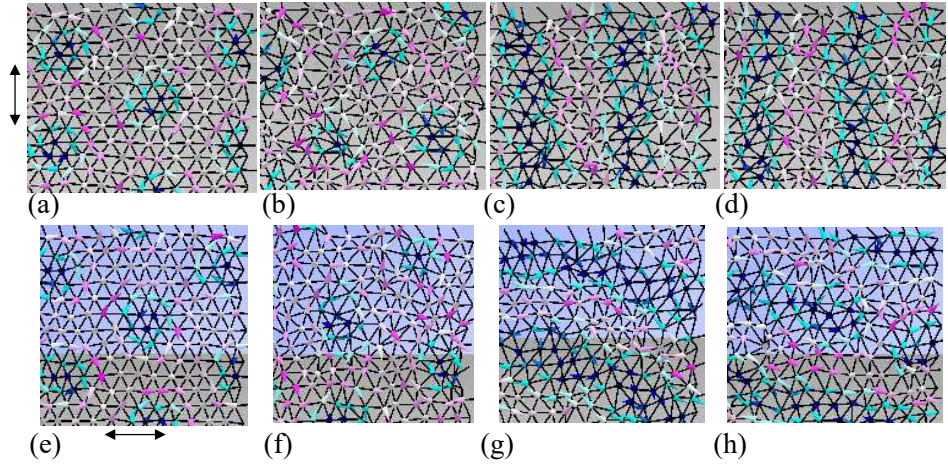}
 \caption{(Color online) Snapshots of lattices with small symbol (cone) for $\sigma$ corresponding to the top-right corner of those of spin configurations in Figs. \ref{fig-10}(a)--\ref{fig-10}(h). An effective stress is applied to the directions ($\updownarrow $) for (a),(b),(c),(d) and ($\leftrightarrow $) for (e),(f),(g),(h).  }
 \label{fig-11}
\end{center}
\end{figure}
It is also interesting to see the lattice structures, although these do not always correspond to those of real materials. The snapshots of lattices are shown by reducing the size of the symbol (cone) for $\sigma$ in Figs. \ref{fig-11}(a)--\ref{fig-11}(h), which correspond to the top-right corner of the spin configurations in Figs. \ref{fig-10}(a)--\ref{fig-10}(h). We find that at low temperature $T\!=\!0.05$ the lattices (Figs.  \ref{fig-11}(a),(e)) are relatively close to the regular lattice as expected,  while the other lattices considerably fluctuate because the temperature $T\!=\!0.3$ is relatively high. These observations indicate that the fluctuation of lattices mainly depends on the temperature. More precisely, the three different straight (=horizontal and two oblique) lines are almost straight and the spaces or separations between them are almost uniform in Figs.  \ref{fig-11}(a),(e), while in the other snapshots these lines deviate from the straight line and the separations are not uniform. We can also see that the separation of the  horizontal line  (oblique lines) in some part of the snapshots in the upper (lower) row becomes slightly narrow. These lattice deformations are due to the imposed uniaxial strains. In addition, combining the observed fact that the lattices are considerably fluctuated at $T\!=\!0.3$ with another fact that the skx phase is stable observed from the variation of $|N_{sk}|$ vs. MCS in Fig. \ref{fig-8}(a), we confirm that the skx phase is stable on fluctuating surfaces. Indeed, we can see from the snapshots in Figs. \ref{fig-11}(b) and \ref{fig-11}(f) in the skx phase,  the lattice structures is considerably fluctuated as described above.

\begin{figure}[t]
\begin{center}
\includegraphics[width=9.5cm]{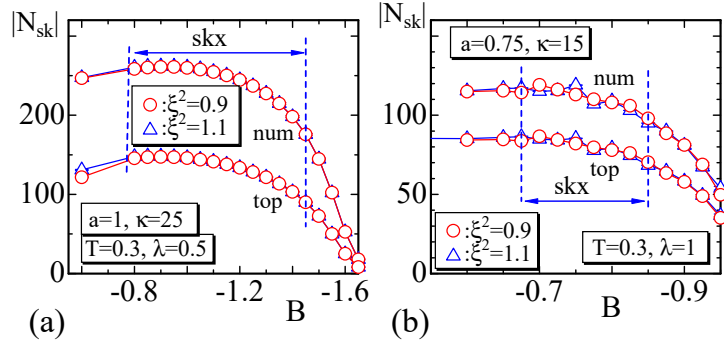}
 \caption{(Color online) The total number of skyrmions $|N_{\rm sk}|$ vs. $B$ under uniaxial strain conditions $\xi^2=0.9$ and  $\xi^2=1.1$ with (a) $a\!=\!1$, $\kappa\!=\!25$, $T\!=\!0.3$, $\lambda\!=\!0.5$, and (b) $a\!=\!0.75$, $\kappa\!=\!15$, $T\!=\!0.3$, $\lambda\!=\!1$. These parameters correspond to those assumed in Fig. \ref{fig-3} and Fig. \ref{fig-4}, respectively. The data plotted in the upper (lower) part of the figures are obtained by the "num" ("top") techniques.
 \label{fig-12}}
\end{center}
\end{figure}
To see the influence of uniaxial strains on the total number $|N_{\rm sk}|$  of skyrmions, we plot $|N_{\rm sk}|$  in Figs. \ref{fig-12} (a),(b) under the conditions corresponding to those of Figs. \ref{fig-5} (c),(d). It is almost clear that there is no influence of uniaxial strains on $|N_{\rm sk}|$, because the results remain unchanged from those of the isotropic case in Figs. \ref{fig-5} (c),(d). Other physical quantities such as $\tau$, $S_0$ and $S_{\rm DM}$ also remain almost unchanged. We should also note that the convergence speed of simulations is almost unchanged from the case of isotropic strains in the previous section. Indeed, the variations of $|N_{\rm sk}|$ vs. MCS  are almost the same as those in Figs. \ref{fig-8}(a),(b). 

\section{Summary and conclusion\label{conclusion}}
In this paper, we study a 2D skyrmion system on fluctuating surfaces with periodic boundary conditions using a Monte Carlo simulation technique. In this model, not only spins but also lattice vertices are integrated into the partition function of the model. From the obtained results, we conclude that skyrmions are stable even on fluctuating surfaces, contrary to initial expectations.

The Dzyaloshinskii-Moriya (DM) interaction is assumed together with the ferromagnetic interaction and Zeeman energy under an external magnetic field. We assume that the lattice fluctuations are only reflected in the DM interaction,  and the magnetoelastic coupling in the ferromagnetic interaction is not taken into consideration. Note that in our model the competition of $S_F$ and $S_{DM}$ suffices to create a skyrmion crystal. To keep the model as simple as possible in order to analyze the elasticity effect, we did not include long-range magnetic dipole-dipole interaction and anisotropy in the model. These ingredients will yield another kind of skyrmions which give rise to another behavior under elasticity and uniaxial strains. 
 In addition to those skyrmion Hamiltonians, the Helfrich-Polyakov Hamiltonian for membranes is assumed.  A surface governed by these Hamiltonians spans a rectangular frame of a fixed projected area, and the surface (or frame) tension is calculated. In this lattice calculation, the effect of stress is introduced by strains, which are imposed on the lattices by fixing the edge lengths of the boundary constant with periodic boundary conditions. In this sense, this effective stress is automatically introduced by the boundary frame.  No mechanical stress can be imposed on the lattice as an input,  but it is obtained as an output of the simulations. 

We obtain numerical results that are consistent with the experimentally-observed fact that a lateral pressure imposed on a 3D material creates skyrmions under a longitudinal magnetic field. Our result indicates that skyrmions are stable on 2D fluctuating surfaces, where there is no crystalline structure in contrast to the rigid lattices including real crystalline materials. 

We further examine how uniaxial strains influence spin configurations. To impose a uniaxial strain on the lattice, we assume an anisotropic condition for the edge length of the rectangular boundary.  
As a result of this calculation, we find that the effective pressure or stress increment makes the stripe domain align to the stress direction under the same magnetic field and temperature as in the case without the uniaxial strain. 
As mentioned in the preceding section, the responses of stripe domains
under uniaxial strains depend on the physical mechanism which generate skyrmions.  Our model is not consistent with the reported experimental results of  Ref. \cite{JDho-etal-APL2003}, but the fact that our stripes are parallel to uniaxial strains may be found in other materials and may have interesting transport applications. Of course, a more general magnetoelastic coupling included not only in $S_{DM}$ but also in $S_F$ should be considered in our future work \cite{Shibata-etal-Natnanotech2015,JWang-etal-PRB2018}.
In contrast, almost no difference is observed in the skyrmion phase, and this also confirms the stability of the skyrmion phase on fluctuating surfaces. 

We should like to comment on the studies of skyrmion stability on fluctuating lattices from the view point of Thiele equation \cite{Thiele-PRL1973}. In Ref.  \cite{Koshibae-Nagaosa-SciRep2017}, an interaction between skyrmion pairs on two separated surfaces is studied by Thiele equation. The authors in Ref.  \cite{Koshibae-Nagaosa-SciRep2017} assume that the bilayer lattice is flat, and hence, effects  of surface fluctuations are not taken into consideration. Therefore, to clarify the influence of the surface fluctuations on the skyrmion stability, it is interesting to study their model on a bilayer which consists of fluctuating and triangulated lattices. One possible technique is simply  to assume that the fluctuating lattices are frozen and only spin variables are time dependent.

Finally, we also comment on a possibility of experimental studies. Skyrmions can only be observed on the materials such as FeGe, MnSi, or ${\rm Cu_2OSeO_3}$ as mentioned in the Introduction, and we have no possibility at present to observe and study skyrmions on elastic surfaces such as ferroelectric polymers under the condition similar to those described in this paper. However, it seems possible to measure  effects of local stresses even on the above-mentioned materials FeGe etc. by using the technique developed for individual manipulation of superconducting magnetic vortex \cite{Kremen-etal-NanoLett2016}. It is interesting to study the stability of skyrmions under local stress or strains. This may be the first step toward the understanding of the role of surface fluctuations in the skyrmion stability.

\noindent
{\bf Acknowledgment}
The authors acknowledge N. Sugita,  Y. Ono and E. Toyoda for the computer simulations and data analyses. The S.E.H. acknowledges Techno AP Co. Ltd.,  Genesis  Co. Ltd., and Kadowaki Sangyo  Co. Ltd. for a financial support for a two-months stay in National Institute of Technology, Ibaraki College, and this stay was also supported in part by JSPS KAKENHI Grant Number JP17K05149. 


\appendix

\section {Graphical Technique to count the skyrmions \label{appendix-A}} 
A graphics software HALCON 13 \cite{graphics-soft} is used to count the skyrmion number $|N_{\rm sk}|$. To use this software efficiently, we produce snapshots with two component colors, red and green, using  the value $\sigma_z (\in [-1,1])$ of spins. The resolution of the graphics of the square area of lattice is approximately $1625\times 1413$, which is the total number of  pixels, for the undeformed lattice ($\Leftrightarrow \xi\!=\!1$ in Eq. (\ref{latt-space-xy})) of the size $N\!=\!10^4$. If the spin configuration is in the skx phase, the central part of skyrmion is colored green and the other part is colored red, and the spins are simply represented by sphere (see Fig. \ref{fig-A1}(a)) in contrast to the snapshots such as Figs. \ref{fig-3}, \ref{fig-4} and \ref{fig-11},  where the spins are represented by cones because of the spin direction. The information of the spin direction is not used to count $|N_{\rm sk}|$ graphically, and for this reason the spins are represented by sphere.
\begin{figure}[t]
\begin{center}
\includegraphics[width=12.5cm]{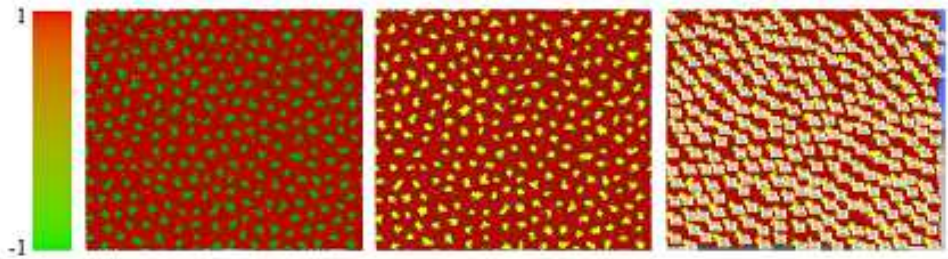}
 \caption{(Color online) (a) Snapshot of a skyrmion configuration obtained under isotropic strain condition for $T\!=\!0.3$, $a\!=\!1$, $\kappa\!=\!25$ and $B\!=\!-1.15$, (b) an output of HALCON 13 \cite{graphics-soft}, where the detected islands are colored yellow, and (c) another output with the numbering of islands. The color is determined from the $z$ component $\sigma_z$ of the variable $\sigma$.  \label{fig-A1}}
\end{center}
\end{figure}

The outline of the graphical measurement is as follows: 
\begin{enumerate}
\item[(1)] In each pixel inside the lattice region, the red and green components of colors are measured and are expressed by two real numbers $(c_{\rm Re},c_{\rm Gr})$, each of which is called "brightness" ranging in $[0,255]$
\item[(2)]  Pixels with $c_{\rm Gr}>c_0$ are selected, where $c_0$ is fixed to  $c_0\!=\!70$   
\item[(3)]  Islands of the selected dots with the area greater than $p_G (=\!300)$ pixels are counted and marked with yellow colors (Fig. \ref{fig-A1}(b)), where this value $p_G$ depends on the resolution of the graphics
\item[(4)]  The marked islands are labeled by numbers, and their area, their center of mass $(x_G,y_G)$ are also obtained (Fig. \ref{fig-A1}(c))   
\end{enumerate}

Due to the assumed periodic boundary condition on the lattices, some of the islands are doubly counted. For this reason, we identify two different islands as a single island if (i) $x_G$'s are almost identical, and the difference of $y_G$'s is almost equal to the side length $L_y$ of the lattice, or (ii) $y_G$'s are almost identical, and the difference of $x_G$'s is almost equal to the side length $L_x$. By this procedure, almost all double counts are removed.  Moreover, the final data are checked by viewing the corresponding snapshots with the eyes.



\end{document}